\documentstyle[aps%
,preprint%
]{revtex}

\def\G{{\cal G}}

\def\row#1,#2,#3{#1 & #2 & #3 \cr}
\def\maplearray#1array([[#2,[#3,[#4])#5{
\pmatrix{\row{#2} \row{#3} \row{#4}}}

\def\tr{\mathop{\rm tr\;}\nolimits}

\def\bl#1,#2{(#1,#2)}

\begin{document}
\title{FINITE ACTION YANG-MILLS SOLUTIONS ON THE GROUP MANIFOLD}
\date{\today}
 \author{T Dereli}
\address{Department of Physics, Middle East Technical University,
06531 Ankara, Turkey, {\tt tdereli{\rm @}rorqual.cc.metu.edu.tr}}

\author{J Schray}
\address{School of Physics and Chemistry, University of Lancaster, Lancs. LA1
4YB, UK, {\tt J.Schray{\rm @}lancaster.ac.uk}}

\author{Robin W Tucker}

\address{School of Physics and Chemistry, University of Lancaster, Lancs. LA1
4YB, UK, {\tt R.Tucker{\rm @}lancaster.ac.uk}}
\maketitle
\begin{abstract}
We demonstrate that the left (and right) invariant Maurer-Cartan forms for any
semi-simple Lie group enable one to construct solutions of the Yang-Mills
equations on the group manifold equipped with the natural Cartan-Killing
metric. For the unitary unimodular groups the Yang-Mills action integral is
finite for such solutions. This is explicitly exhibited for the case of
$SU(3)$.
\end{abstract}

\section{Introduction}
Classical solutions of the Yang-Mills field equations have proved to be of
immense value in the
development of both mathematical and physical theory.
Instanton and monopole solutions have provided a fertile source of
approximation techniques in quantum field theory  and recent developments
in duality theory have led to exciting new computational schemes in
differential geometry based on supersymmetric Yang-Mills actions.
Although self-duality has featured prominently in the construction of
finite-action solutions not all such field configurations fall into this
category. One of the earliest  solutions found in $SU(2)$ 
Yang-Mills theory was the ``meron''
\cite{Alfaro-YM,Alfaro-O4xO2,Alfaro-GT,Glimm,Cervero,Howe,Minzoni}.
Its generalisations 
 gave rise to an interesting proposal for the  confinement of quarks
\cite{Callan,Glimm-QC}.  Little attention has been devoted to the exploration
of finite action solutions associated with higher rank Lie groups
in four \cite{Chakrabarti}  or higher dimensions \cite{Popov}.

It is the purpose of this note to point out that solutions of ``meron'' type
can occur in a much broader context. We demonstrate below that there exist
similar finite-action solutions of the Yang-Mills field equations associated
with {\it any compact semi-simple Lie group} if the action functional for the
field
equations is constructed over the group manifold itself in terms of a group
invariant metric on the Lie algebra.  Of particular interest for physics are
the unitary unimodular groups.  Such a broad class of solutions may have some
relevance to one or more of the many compactification schemes that feature in
the dimensional reduction of higher dimensional descriptions of the
fundamental interactions.  The explicit construction of any solution in this
class depends upon the manner in which the group manifold is coordinated. We
content ourselves here with the explicit construction of a finite-action
solution associated with $SU(3)$.

\section{Yang-Mills solutions}

Suppose $G$ is a semi-simple Lie group.
Let $\{\omega^{\alpha}\}$ be the set of left invariant
basis 1-forms on  $G$ that satisfy
the Maurer-Cartan structure equations
\begin{equation}
 d\omega^{\alpha}= -\frac{1}{2} C_{\beta \gamma}{}{}^{\alpha}
\omega^{\beta} \wedge \omega^{\gamma} 
\label{MC}
\end{equation}
where $C_{\beta \gamma}{}{}^{\alpha}$ are the structure
constants of the Lie algebra  of $G$. If we define the dual basis
of vectors $\{X_{\alpha}\}$ so that $\omega^{\alpha}(X_{\beta})
=\delta^{\alpha}{}_{\beta}$ then 
\begin{equation}
[ X_{\alpha} , X_{\beta} ] = C_{\alpha \beta}{}{}^{ \gamma} X_{\gamma}. 
\end{equation}
Let $g$ denote the Cartan-Killing metric on the group manifold $G$
so that
\begin{equation}
g_{\alpha \beta} \equiv g(X_{\alpha},X_{\beta})
= C_{\alpha \gamma}{}{}^{\delta} C_{\beta \delta}{}{}^{\gamma}. 
\label{Cartan-Killing}
\end{equation}
We also introduce 
an irreducible representation%
\footnote{ For $G=SU(N)$ these matrices would be 
a set of anti-Hermitian $N \times N$ traceless matrices corresponding to
the fundamental representation.}
 of the Lie algebra of $G$ by
$N \times N$   
matrices  $\{\lambda_\alpha \}$ 
 that satisfy
\begin{equation}
 [ \lambda_{\alpha} , \lambda_{\beta} ]= C_{\alpha \beta}{}{}^{ \gamma} 
\lambda_{\gamma}. 
\end{equation}
Thus the matrices are scaled in such a way that
\begin{equation}
 \tr({\rm ad} \,\lambda_{\alpha}\,\, {\rm ad} \,\lambda_{\beta}) 
= g_{\alpha \beta}. 
\end{equation}
Define a Yang-Mills potential 1-form
\begin{equation}
 A = c \,\, \omega^{\alpha}  \lambda_{\alpha} 
\end{equation}
where c is a constant to be determined.
The corresponding gauge field 2-form is
\begin{equation}
 F \equiv dA + \frac{1}{2} [A , A] = - c(c-1) d\omega^{\alpha}
\lambda_{\alpha}.
\end{equation}
Now consider the Hodge dual  $\ast F$ with respect to the
Cartan-Killing metric
and substitute it in the Yang-Mills equation
\begin{equation}
 d \ast F + A \wedge \ast F - \ast F \wedge A = 0. 
\end{equation}
We find
\begin{equation}
d \ast F = -c \, (c-1) \, d \ast d \omega^\alpha \lambda_\alpha
\end{equation}
and
\begin{equation}
A \wedge \ast F - \ast F \wedge A = c^2 (c-1) \, {C_{\beta \gamma}}^\alpha
\omega^\beta \wedge \ast d \omega^\gamma \lambda_\alpha.
\end{equation}
Excluding the trivial solutions $c= 0,1$ the Yang-Mills equation is satisfied 
provided we have
\begin{equation}
 d \ast d \omega^{\alpha} = -c \, C_{\beta \gamma}{}{}^{\alpha}
\omega^{\beta} \wedge \ast d \omega^{\gamma}. 
\label{YM}
\end{equation}
From Eq.~(\ref{MC}) the left-hand side is
\begin{eqnarray}
d \ast d \omega^{\alpha} &=& -{1\over 2} {C_{\beta \gamma}}^{\alpha}
d \ast (\omega^\beta \wedge \omega^\gamma)\nonumber\\
&=& -{1\over 2} {C_{\beta \gamma}}^{\alpha} d 
\left({1\over (N -2)!} \, {\epsilon^{\beta\gamma}}_{\delta\rho\ldots}
\, \omega^\delta \wedge \omega^\rho \wedge \ldots \right),
\end{eqnarray}
where ${\epsilon^{\beta\gamma}}_{\delta\rho\ldots}$ is the Levi-Civita
alternating density associated with the Cartan-Killing metric.
With Eqs.~(\ref{MC}) and (\ref{Cartan-Killing}) it follows that
\begin{eqnarray}
d \ast d \omega^\alpha &=& - {1\over 2} {C_{\beta\gamma}}^\alpha
d \omega^\delta \wedge \ast (\omega^\beta \wedge \omega^\gamma \wedge
\omega_\delta)\nonumber\\
&=&{1\over 4} {C_{\beta\gamma}}^\alpha {C_{\rho\sigma}}^\delta \omega^\rho
\wedge \omega^\sigma \wedge \ast (\omega^\beta \wedge \omega^\gamma \wedge
\omega_\delta).
\end{eqnarray}
With the aid of the identities
\begin{eqnarray}
\omega^{\alpha} \wedge \ast (\omega^{\beta} \wedge \omega^{\gamma} \wedge
\omega^{\delta}) &=&
g^{\alpha \beta} \ast (\omega^{\gamma} \wedge \omega^{\delta})
-g^{\alpha \gamma} \ast (\omega^{\beta} \wedge \omega^{\delta}) 
+g^{\alpha \delta} \ast (\omega^{\beta} \wedge \omega^{\gamma}),
\label{eq11}\\
\omega^{\alpha} \wedge \ast (\omega^{\beta} \wedge \omega^{\gamma}) &=&
-g^{\alpha \beta} \ast \omega^{\gamma}
+g^{\alpha \gamma} \ast \omega^{\beta}
\label{eq12}
\end{eqnarray}
this gives
\begin{equation}
d \ast d \omega^\alpha = {1\over 2} \, g^{\beta\rho} g^{\gamma\sigma}
{C_{\beta\gamma}}^\alpha {C_{\rho\sigma}}^\delta \ast \omega_\delta.
\end{equation}
On the other hand the right-hand side of Eq.~(\ref{YM}) may be written
\begin{eqnarray}
-c \, C_{\beta \gamma}{}{}^{\alpha} \omega^{\beta} \wedge
\ast d \omega^{\gamma}
&=&  {c\over 2} \, {C_{\beta\gamma}}^\alpha \omega^\beta \wedge
{C_{\rho\sigma}}^\gamma \ast ( \omega^\rho \wedge \omega^\sigma)
\qquad\mbox{(from Eq.~(\ref{MC}))}
\nonumber\\
&=&-c \, {C_{\beta\gamma}}^\alpha {C_{\rho\sigma}}^\gamma g^{\beta\rho} \ast
\omega^\sigma
\qquad\mbox{(by Eq.~(\ref{eq12}))},
\end{eqnarray}
with $g^{\alpha \beta}$ being the inverse of the Cartan-Killing metric.
Hence the Yang-Mills equation is satisfied if
\begin{equation}
{1\over 2} {C_{\beta\gamma}}^\alpha C^{\beta\gamma\delta} = -c \,
{C_{\beta\gamma}}^\alpha C^{\beta\delta\gamma},
\end{equation}
where 
\begin{equation}
 C_{\alpha \beta \gamma} \equiv C_{\alpha \beta}{}{}^{\delta}
g_{\delta \gamma} 
\end{equation}
are totally skew-symmetric.
This implies $c = {1\over 2}$.

If we now restrict to the compact unimodular unitary groups, the group
manifold has a finite volume $\int_G \ast 1$ and the above solution yields
a finite  Yang-Mills action: 
\begin{equation}
 I[A] = \int_{G} \tr(F \wedge \ast F ) = 
\frac{ C_{\alpha \beta \gamma} C^{\alpha \beta \gamma}}{32} 
\int_{G} \ast 1,
\label{YMaction}
\end{equation}
where we have used Eq.~(\ref{eq12}) and
\begin{equation}
\omega^{\alpha} \wedge \ast \omega^{\beta} = g^{\alpha \beta} \ast 1.
\end{equation}

The construction of group volumes is not trivial in general
\cite{Marinov,Murnaghan,Beg,Moshinsky,Nelson,Sudbery,Holland} and the precise
values depend on the choice of normalisation of the group generators.  We
illustrate this by computing the above action for $G = SU(3)$.

If we denote the anti-Hermitian generators of the Lie algebra of $SU(3)$ by 
\begin{eqnarray}
&\lambda_1 = {\sqrt{2} \over 2} \, 
 \pmatrix{ 0 & i & 0 \cr i & 0 & 0 \cr 0 & 0 & 0 \cr },
\lambda_2 = {\sqrt{2} \over 2} \, 
 \pmatrix{ \phantom{-}0 & 1 & 0 \cr -1 & 0 & 0 \cr
  \phantom{-}0 & 0 & 0 \cr },
\lambda_3 = {\sqrt{2} \over 2} \, 
 \pmatrix{ i & \phantom{-}0 & 0 \cr 0 & -i & 0 \cr
  0 & \phantom{-}0 & 0 \cr},\nonumber\\
&\lambda_4 = {\sqrt{2} \over 2} \, 
 \pmatrix{ 0 & 0 & i \cr 0 & 0 & 0 \cr i & 0 & 0 \cr},
\lambda_5 = {\sqrt{2} \over 2} \, 
 \pmatrix{ \phantom{-}0 & 0 & 1 \cr \phantom{-}0 & 0 & 0 \cr
  -1 & 0 & 0 \cr },
\lambda_6 = {\sqrt{2} \over 2} \,
 \pmatrix{ 0 & 0 & 0 \cr 0 & 0 & i \cr 0 & i & 0 \cr },\nonumber\\
&\lambda_7 = {\sqrt{2} \over 2} \, 
 \pmatrix{ 0 & \phantom{-}0 & 0 \cr 0 & \phantom{-}0 &
  1 \cr 0 & -1 & 0 \cr },
\lambda_8 = {\sqrt{6} \over 6} \, 
 \pmatrix{ i & 0 & \phantom{-}0 \cr 0 & i & \phantom{-}0 \cr
  0 & 0 & -2 \, i \cr}
\end{eqnarray}
which are just ${i \over \sqrt{2}}$ times the Gell-Mann ``$\lambda$-matrices''
normalised such that $\tr (\lambda_i \lambda_j) = -\delta_{ij}$,
then the group parametrisation of $SU(3)$ given by Holland \cite{Holland}
reads
\begin{equation}
\G(\bbox \theta) = 
e^{- \sqrt{2} \, \theta^1 \lambda_3}
e^{- \sqrt{2} \, \theta^2 \lambda_2}
e^{- \sqrt{2} \, \theta^3 \lambda_3}
\;
e^{- \sqrt{6} \, \theta^4 \lambda_8}
e^{- \sqrt{2} \, \theta^5 \lambda_4}
\;
e^{- \sqrt{2} \, \theta^6 \lambda_3}
e^{- \sqrt{2} \, \theta^7 \lambda_2}
e^{- \sqrt{2} \, \theta^8 \lambda_3}
\end{equation}
The parameter ranges for the eight angles are $0 \leq \theta^1,
\theta^3, \theta^6, \theta^8 \leq 2 \, \pi$, $0 \leq \theta^4
\leq \pi$, and $0 \leq \theta^2, \theta^5, \theta^7 \leq {1 \over 2} \pi$.

The Maurer-Cartan 1-forms $\omega^1$, $\omega^2$, \ldots,
$\omega^8$ are defined by
\begin{equation}
\omega^i = - \tr (\lambda_i \G^{-1} d\G).
\end{equation}
The Cartan-Killing metric may be written in this parametrisation as 
\begin{equation}
-g_{CK} = \omega^1 \otimes \omega^1 + \omega^2 \otimes \omega^2
+ \ldots + \omega^8 \otimes \omega^8\nonumber
\end{equation}
and the wedge product of the Cartan-Killing 1-forms, $\ast 1$, is calculated
to be
\begin{equation}
\ast 1 = \omega^1 \wedge \omega^2 \wedge \ldots \wedge \omega^8
= 8 \, \sqrt{3} \, \sin(2 \, \theta^2) \, \sin(2 \, \theta^5) \,
\sin^2(\theta^5) \, \sin(2 \, \theta^7) \,
d\theta^1 \wedge d\theta^2 \wedge \ldots \wedge d\theta^8.
\end{equation}
We may integrate this to obtain the group volume $V_{SU(3)}$ of $SU(3)$ with
respect to the Cartan-Killing metric volume element:
\begin{eqnarray}
V_{SU(3)} &=& \int_{SU(3)} \! *1 \nonumber\\&=&
\int_0^{2 \, \pi} \!\!\!\! \int_0^{\pi \over 2} \!\!\! \int_0^{2 \, \pi}
\!\!\!\! \int_0^{\pi} \!\!\! \int_0^{\pi \over 2} \!\!\! \int_0^{2 \, \pi}
\!\!\!\! \int_0^{\pi \over 2} \!\!\! \int_0^{2 \, \pi}
8 \, \sqrt{3} \, \sin(2 \, \theta^2) \, \sin(2 \, \theta^5) \,
\sin^2(\theta^5) \, \sin(2 \, \theta^7)
\nonumber\\&&{}\times
 d\theta^1 \, d\theta^2 \, d\theta^3 \, d\theta^4 \, d\theta^5 \, d\theta^6
\, d\theta^7 \, d\theta^8\nonumber\\&=&
8 \, \sqrt{3} \,\left[ (2 \,\pi) \times 1 \times (2 \,\pi)
\times \pi \times {1\over 2} \times (2 \,\pi) \times 1 \times (2 \,\pi)
\right] = 64 \, \sqrt{3} \, \pi^5
\label{volume}
\end{eqnarray}
Eqs.~(\ref{YMaction}), and (\ref{volume}) give explicit formulae for the
Yang-Mills action associated with the ``meron'' type solution discussed
above.
For $SU(3)$, this solution has both Pontryagin numbers zero.


\section{Conclusion}

We have shown that for any semi-simple Lie group with general element $\G$ the
potential form $A=c \, 
\G^{-1}\, d\G$ is a solution of the Yang-Mills equation on a group manifold
with
the natural Cartan-Killing metric, if c takes the value ${1\over 2}$. By a
simple local gauge transformation $A=(1-c) \, \G\, d\G^{-1}$ is an equivalent
solution. If the group is unimodular, both left and right invariant group
volume elements can be chosen to coincide, so for compact groups the
Yang-Mills action is finite for such solutions.  For $SU(3)$ the result is
$$\vert I[A]\vert=\mu^2 96 \sqrt{3} \pi^5$$ for the metric $g=\mu\, g_{CK}$
where $\mu$ is any real constant.

We have also briefly examined the more general ansatz 
$$A=\sum_{i=1}^{8} c_i\, \omega^i\,\lambda_i.$$
For a constant vector
$\{c_i\}$,
$\sum_{i=1}^{8} c_i \omega^i$ is the most general left 
invariant 1-form on the group. One
finds that $A$ generates a solution provided the components
$\{c_i\}$ satisfy a set of coupled quartic equations. For $SU(3)$ we
demand that these equations have real solutions. Such solutions exist but
may be shown to be related to the solution discussed above by  constant
gauge transformations (automorphisms of the Lie algebra). The remaining
 complex solutions  give rise to finite but complex
Yang-Mills actions.

There have been a number of papers
\cite{Tchrakian,Sac,O'Brien}
devoted to the generalisations of the Yang-Mills solutions to higher
dimensions. In their search for the higher dimensional analogues of
instantons and merons these papers have modified the form of 
the original Yang-Mills
action in an attempt to preserve the conformal properties that the theory
exhibits in four dimensions.
In this note  we have given a simple proof of the existence of finite
action solutions to the original Yang-Mills equation expressed in terms of
the natural group invariant metric provided by the group manifold itself.
 
\acknowledgements

RWT and TD are grateful to NATO grant number CRG-941210 which made this
work possible.  RWT is also 
grateful to the Human Capital and Mobility Programme of the European
Union for partial support.  TD would like to thank M S Marinov for useful
discussions.

\end{document}